\documentclass[preprint2]{aastex}

\usepackage{amsmath}
\usepackage{xspace} 
\usepackage[utf8]{inputenc}

\renewcommand{\deg}{$\,^{\circ}$\xspace}

\newcommand{\MQ}{$M/Q$\xspace}	
\newcommand{\Sval}{$S$\xspace}	
\newcommand{\Bval}{$B$\xspace}	

\usepackage{multirow}
\def\Put(#1,#2)#3{\leavevmode\makebox(0,0){\put(#1,#2){#3}}}

\usepackage{natbib}
\shorttitle{Temporal variation of SEP ratios}
\shortauthors{Zelina et al.}

\begin{document}

\title{Time Evolution of Elemental Ratios in Solar Energetic Particle events}

\author{P. Zelina and S. Dalla}
\affil{University of Central Lancashire, Preston PR1 2HE, UK}
\email{pzelina@uclan.ac.uk}

\and

\author{C. M. S. Cohen and R. A. Mewaldt}
\affil{California Institute of Technology, Pasadena, CA 91125, USA}

\begin{abstract}

Heavy ion ratio abundances in Solar Energetic Particle (SEP) events, e.g.~Fe/O, often exhibit decreases over time. Using particle instruments on the ACE, SOHO and STEREO spacecraft, we analysed heavy ion data from 4 SEP events taking place between December 2006 and December 2014. We constructed 36 different ionic pairs and studied their time evolution in each event. We quantified the temporal behaviour of abundant SEP ratios by fitting the data to derive a decay time constant $B$. We also considered the ratio of ionic mass--to--charge for each pair, the $S$ value given e.g.~for Fe/O by $S_{\rm Fe/O} = (M/Q)_{\rm Fe}\big/(M/Q)_{\rm O}$. We found that the temporal behaviour of SEP ratios is ordered by the value of $S$: ratios with $S>1$ showed decreases over time (i.e.~$B<0$) and those with $S<1$ showed increases ($B>0$). We plotted $B$ as a function of \Sval and observed a clear monotonic dependence: ratios with a large $S$ decayed at a higher rate. A prominent discontinuity at $S=2.0$ (corresponding to He/H) was found in 3 of the 4 events, suggesting anomalous behaviour of protons. The X/H ratios often show an initial increase followed by a decrease, and decay at a slower rate. We discuss possible causes of the observed $B$ versus \Sval trends within current understanding of SEP propagation.

\end{abstract}


\section{Introduction}
\label{sec:intro}

Solar energetic particles (SEPs) are ions and electrons released into the interplanetary medium due to solar eruptive activity. They are accelerated by solar flares and coronal mass ejections (CMEs) and can be observed as particle intensity increases by instruments onboard spacecraft. SEPs (mainly H, He, electrons) can be detected at locations widely separated from the parent eruptive event in longitude and latitude (e.g.~\citealt{gomez2015,dalla2003_grl}). In large, so called gradual SEP events, mean ionic abundances are similar to the solar wind ones but event--to--event variation of ratios such as Fe/O can vary by significantly more than typical solar wind ratios. Type II radio bursts from the associated CME--driven shock are observed in addition to type III radio bursts generated by particles released from solar flares. The wide longitudinal extent of SEP events is usually ascribed to acceleration at extended regions, such as shocks, because according to the accepted paradigm for SEP propagation, SEPs travel along magnetic field lines without significant transport perpendicular to the average interplanetary magnetic field \citep{reames1999}. SEP events with high Fe/O abundance and rich in $^3$He are called impulsive events. These events are of short duration and commonly associated only with type III radio bursts. According to the standard paradigm, impulsive SEPs should be detectable only in narrow regions of $<\!20\,^{\circ}$ in longitude, i.e.~locations with good magnetic connection to the flare.

Since the launch of the STEREO spacecraft, there have been many reports of simultaneous observations by multiple spacecraft that showed that proton and electron SEP events can indeed be detected at locations widely separated in longitude (e.g.~\citealt{dresing2012,lario2013,dresing2014,richardson2014,gomez2015}). In a study of $^3$He rich events detected by multiple spacecraft, \citet{wiedenbeck2013} reported an impulsive event that was observed by three spacecraft, two of which were separated by 136\deg, but as part of a statistical study of 17 impulsive events they showed that simultaneous SEP detections at two spacecraft at longitudinal separation $>\!60\,^{\circ}$ were not uncommon. An SEP event rich in Fe was reported by \citet{cohen2014}, where two spacecraft separated by 135\deg in longitude detected enhanced average Fe/O values, which showed a dependence on longitude. In separate studies, \citet{gomez2015} and \citet{zelina2015_icrc} used data from 3 spacecraft and found that Fe and O SEPs from a single parent active region can reach locations widely separated in longitude and be detected over almost 360\deg.

SEP ratios, such as Fe/O, often decrease over the duration of an SEP event. Using 3--hour averaged data, \citet{scholer1978} observed Fe/O ratios decreasing in time while the C/O ratios were time independent. \citet{tylka1999} reported observations of the temporal evolution of several heavy ion ratios during the 1998 April 20 SEP event that were ordered by their \MQ values, where $M$ is atomic mass of an SEP ion and $Q$ its charge. \citet{mason2012} studied the temporal evolution of Fe/O, O/He, and He/H ratios in 17 SEP events, where in majority of events the ratios exhibited temporal variation. \citet{tylka2013} used Ulysses and near--Earth spacecraft data to show that Fe SEPs can reach high heliographic latitudes. They also observed the characteristic Fe/O decrease over time at the two spacecraft, both of which had poor magnetic connection to the parent flare. In all the considered events analysed by \citet{zelina2015_icrc} the Fe/O ratio decreased over time, a behaviour therefore identified as a common feature of SEP events.

Several researchers have proposed interpretations of the observed time dependence of elemental ratios, either as an effect of acceleration or of transport. It was suggested that the high Fe/O ratio early in the SEP event is a result of an initial flare component (rich in Fe) while the decrease that follows is associated with a shock--accelerated component (with lower Fe/O) later in the event \citep{cane2003}. \citet{tylka1999} explained it as due to the ions with high \MQ (i.e.~Fe) escaping the accelerating shock region more easily. Others proposed that the observed time dependence is a propagation effect due to the rigidity dependence of the mean free path \citep{scholer1978,mason2012}, or more generally transport effects \citep{tylka2013}.

The observed temporal variation of SEP ratios may be related to \MQ--dependent cross--field transport of SEPs, where SEPs with different \MQ follow different trajectories. Gradient and curvature drifts in the Parker spiral magnetic field depend on \MQ and kinetic energy \citep{dalla2013}. Full--orbit simulations of heavy ions show that drifts are an important mechanism of perpendicular transport that can distribute SEPs across the interplanetary magnetic field \citep{marsh2013}. The charge state affects how much a particle can drift across the mean magnetic field, since particles with higher \MQ exhibit more drift. Simulations by \citet{dalla2015_icrc,dalla2016} show that an SEP model including drifts can qualitatively reproduce the decrease in time of the Fe/O ratio.

In this work, we studied the temporal behaviour of SEP abundance ratios for 4 SEP events in a quantitative manner. We systematically analysed the decay in time of a number of elemental ratios, including Fe/O and less commonly used ratios such as Fe/C, Fe/Mg, and Fe/Si. We quantified the temporal evolution of SEP ratios by fitting the time profiles to an exponential function and deriving a decay time constant, and studied any correlations of this parameter with ionic \MQ values of the 2 species in the ratio. Section~\ref{sec:observations} presents SEP data for the selected events, followed by a Discussion (Section~\ref{sec:discussion}) and Conclusions (Section~\ref{sec:conclusions}).

\section{Observations}
\label{sec:observations}

\begin{table*}[thb]
\begin{center}
\caption{Details of solar flares and coronal mass ejections (CMEs) obtained from {\tt SolarMonitor.org} and the CDAW CME catalogue. The position of the backside flare (event \#3) was calculated using STEREO FITS files. The flare class for this event was estimated by \citet{pesce2015}. \label{tab:solar_events}}

{
\begin{tabular}{llllll} 
\tableline\tableline
 
    &	\multicolumn{4}{l}{Event} \\

				&	\#1 &		\#2 &		\#3 &		\#4 \\

\tableline

Year				&	2012 		& 2006 			& 2014 			& 2014 \\
Flare start time	&	Aug 31/19:45 & Dec 13/02:14	& Sep 01/11:00	& Feb 25/00:39 \\
Flare class			&	C8			& X3.4 			& X2.1 			& X4.9 \\
Solar disk location	&	S16E42 		& S07W22 		& N14E129 		& S12E77 \\
CME start time		&	22:00 		& 02:54 		& 11:12 		& 01:25 \\
CME speed [km/s]	& 	1442 		& 1774 			& 1901 			& 2147 \\
 
\tableline
 
\end{tabular}
}

\end{center}
\end{table*}

\begin{table*}[thb]
\begin{center}
\caption{Details of the SEP events. The duration is the time span with good Fe count statistics in {\em day of year} units, for which are heavy ion ratios plotted and analysed. $\Delta\phi$ is longitudinal separation (positive is flare west of the spacecraft footpoint), $\Delta\theta$ is latitudinal separation (positive is flare north of the spacecraft), $A_2/A_1$ is the ratio of Fe/O values (final/initial), $\Delta t$ the time over which the Fe/O decrease occurs, and $B$ is the derived exponential decay time constant. \label{tab:sep_events}}
 
{
\begin{tabular}{llllll} 
\tableline\tableline
 
    &	\multicolumn{4}{l}{Event} \\

				&	\#1 &	\#2 &	\#3 &	\#4 \\

\tableline

Year		&			2012 &	2006 &	2014 &	2014 \\
Duration [DOY]	&		244--248 &	347--349 &	244--248 &	56--60 \\	
Best connected s/c	&	STB &	SOHO,ACE &	STB &	STB \\
SW speed [km/s]	&	325 &	650 &	450 &	500 \\		
$\Delta\phi$	&		-5\deg &	-17\deg &	-25\deg &	32\deg \\
$\Delta\theta$	&		-12\deg &	-6\deg &	21\deg &	-19\deg \\
Fe/O $A_2/A_1$	&		0.030 &	0.121 &	0.027 &	0.150 \\
Fe/O $\Delta t$ [day] &	1.25 &	1.25 &	1.25 &	2.75 \\
Fe/O $B$ [$\rm day^{-1}$]	&	$-0.92\pm0.06$ &	$-0.54\pm0.06$ &	$-1.1\pm0.1$ &	$-0.01\pm0.02$ \\
 
\tableline
 
\end{tabular}
}

\end{center}
\end{table*}

\begin{table}[htb]
\begin{center}

\caption{Atomic mass number $M$, ionic charge state number $Q$ and \MQ of abundant SEP ions. The ionic charge states are from \citet{luhn1985} except for H and He, which are assumed to be fully ionised. The heavy ion charge state values are subject to 5\% systematic uncertainty. \label{tab:charge_states}}

\begin{tabular}{lrrc}

\tableline\tableline

Element	& \multicolumn{1}{c}{$M$}	& \multicolumn{1}{c}{$Q$}	& \multicolumn{1}{c}{\MQ} \\

\tableline

H	& 1		& 1.00		& 1.00 \\
He	& 4		& 2.00		& 2.00 \\
C	& 12	& 5.70	 	& 2.11 \\
N	& 14	& 6.37 		& 2.20 \\
O	& 16	& 7.00		& 2.29 \\
Ne	& 20	& 9.05		& 2.21 \\
Mg	& 24	& 10.70 	& 2.24 \\
Si	& 28	& 11.00 	& 2.55 \\
Fe	& 56	& 14.90 	& 3.76 \\

\tableline

\end{tabular}
\end{center}
\end{table}

\begin{table*}[thb]
\begin{center}

\caption{Table of $S$ values, for all common pairs of abundant SEP elements. $ \rm Ratio = \frac{Element (2)}{Element (1)}$. \label{tab:s_values}}

{ 
\begin{tabular}{lcccccccc}

\tableline\tableline

Element & \multicolumn{7}{c}{Element (2)} \\

\multicolumn{1}{c}{(1)} & He & C & N & O & Ne & Mg & Si & Fe \\

\tableline

H	& 2.00	& 2.11	& 2.20	& 2.29	& 2.21	& 2.24	& 2.55	& 3.76 \\
He	& 1		& 1.05	& 1.10	& 1.14	& 1.10	& 1.12	& 1.27	& 1.88 \\
C	& 		& 1		& 1.04	& 1.09	& 1.05	& 1.07	& 1.21	& 1.79 \\
N	& 		& 		& 1		& 1.04	& 1.01	& 1.02	& 1.16	& 1.71 \\
O	& 		& 0.92	& 0.96	& 1		& 0.97	& 0.98	& 1.11	& 1.64 \\
Ne	& 		& 		& 		& 		& 1		& 1.01	& 1.15	& 1.70 \\
Mg	& 		& 		& 		& 		& 		& 1		& 1.13	& 1.68 \\
Si	& 		& 		& 		& 		& 		& 		& 1		& 1.48 \\

\tableline

\end{tabular}

 }
\end{center}
\end{table*}

For the study of heavy ion particles, we used SEP data measured in situ by the following particle instruments: Solar Isotope Spectrometer (SIS; \citealt{stone1998}) onboard Advanced Composition Explorer (ACE), Energetic and Relativistic Nuclei and Electron (ERNE; \citealt{torsti1995}) onboard SOlar and Heliospheric Observatory (SOHO), and Low Energy Telescopes (LET; \citealt{mewaldt2008}), and High Energy Telescopes (HET; \cite{rosenvinge2008}) onboard the Solar Terrestrial Relations Observatory (STEREO) Ahead (STA) and Behind (STB) spacecraft.

We examined SEP events between December 2006 and December 2014 with the Fe particle intensity above $10^{-3}\:\rm (cm^2\,s\,sr\,MeV/nuc)^{-1}$ in the $10-12\:\rm MeV/nuc$ energy channel for STEREO and $10^{-4}\:\rm (cm^2\,s\,sr\,MeV/nuc)^{-1}$ in the $10.7-15.8\:\rm MeV/nuc$ energy channel for ACE. The lower threshold in the ACE data is due to $\sim\!10\rm x$ larger geometrical factor of the ACE/SIS instrument compared to the STEREO/LET instruments. We selected 4 SEP events: 2006 December 13, 2012 August 31, 2014 February 25 and 2014 September 1, which have intensity profiles with monotonic rise and decay phases. Each of the selected events was linked to a single parent active region. Details of the solar eruptive events are given in Table~\ref{tab:solar_events}.

Details of the SEP events are given in Table~\ref{tab:sep_events}. For each SEP event we used the spacecraft with best magnetic connection to the flare. In all events other spacecraft did not measure sufficient particle intensities of heavy ion SEPs to quantitatively analyse the observed temporal variation of heavy ion ratios, or intensity time profiles did not show monotonic rise and decay phase. In Table~\ref{tab:sep_events} the duration of an SEP event indicates the number of days when there were sufficient Fe particles to yield good statistics. We used the local solar wind speed value at the beginning of an SEP event at the spacecraft to calculate the nominal Parker spiral footpoint of the spacecraft on the solar surface $\phi_{sc}$. The angular separation in longitude between the flare and the spacecraft is calculated as $\Delta\phi = \phi_{flare} - \phi_{sc}$, where $\phi_{flare}$ is the longitude of the flare. A positive $\Delta\phi$ means that the flare is western with respect to the observer footpoint. In a similar manner, latitudinal separation between the flare and the observer was calculated as $\Delta\theta = \theta_{flare} - \theta_{sc}$. A positive $\Delta\theta$ means the flare is north of the spacecraft.

For each event, time intensity profiles for all abundant elements, which include H, He, C, N, O, Ne, Mg, Si and Fe, were analysed and ionic ratios constructed. For ratios observed by STEREO/LET, we used the lowest energy channel common for all elements that is $4.0-4.5$~MeV/nuc. ACE/SIS energy channels do not cover exactly the same energy range as STEREO/LET for any of the heavy ions. Therefore, we used the closest available channels to the STEREO/LET energy channel. Note that the heavy ion energy bins in ACE/SIS data change depending on the analysed element due to the SIS instrument response function for the analysed nuclei \citep[see][Figure~19 for reference]{stone1998}. For example, the lowest energy bin for He is $3.4-4.7\:\rm MeV/nuc$ while the lowest Fe energy bin is $10.7-15.8\:\rm MeV/nuc$. The energy channels for a pair of two elements were chosen to be the closest to each other, e.g.~in order to obtain the Fe/O ratio, Fe intensity at $10.7-15.8\:\rm MeV/nuc$ was divided by O intensity measured at $10.0-13.1\:\rm MeV/nuc$. We used proton measurements from SOHO/ERNE to complement the heavy ion data by ACE/SIS. For all the intensities, uncertainties were calculated as $\sigma = \sqrt{ N }$, where $N$ is the particle count within the accumulation time.

For each pair of SEP elements X$_1$ and X$_2$ we define a parameter \Sval given by

\begin{equation}
S_{ {\rm X}_2 / {\rm X}_1 } \equiv \left( \frac{M_2}{Q_2} \right) \Bigg/ \left(\frac{M_1}{Q_1}\right)
\end{equation}

\noindent where $M_i$ is atomic mass number and $Q_i$ is the ionic charge state in elementary charge units, for species X$_i$.

SEPs in the interplanetary medium are partially ionised. Measurement of the charge state $Q$ is challenging and is not routinely carried out for all SEP events \citep{klecker2006}. For this reason, for the purpose of calculating the $S$--values, we use charge state measurements by \citet{luhn1985}, shown in Table~\ref{tab:charge_states}, which provide an estimate of ionic charge states at $\approx\!0.5-3.3\:\rm MeV/nuc$ in a similar energy range to the one considered in our study. The charge state measurements given in \citet{luhn1985}, obtained by averaging over a number of SEP events, are subject to systematic errors of 5\%. {
It should be noted that $Q$ can vary event to event and in some SEP events it increases with energy \citep{klecker2006,mewaldt2006}. }

{
For the event of 2006 December 13, one of the events in our analysis, we estimate from the SAMPEX data, using the same method described by \citet{oetliker1997}, that the iron charge state was $Q_{\rm Fe} = 16.2\substack{+1.7 \\ -1.5}$ at $25-90\:\rm MeV/nuc$. Therefore we can say that for this event, the value of $Q$ at the lower energies that we consider in our study, as given in Table~\ref{tab:charge_states}, is not inconsistent with the measured value at higher energies. }

{
To our best knowledge there were no operational instruments measuring ionic charges in the other 3 events. Charge states are known to vary event to event and the charge state of iron in particular can take a broad range of values $Q_{\rm Fe} \approx 10-20$ \citep{labrador2005}. Other ionic charge states vary too but to a lesser extent.}

We use the atomic mass number of the dominant isotope as a mass estimate for an SEP species. The difference between this value and the isotopic SEP compositions, e.g.~given by \citet{anders1989}, is estimated to be $\leq2\%$, but it is substantially less than the uncertainty in SEP charge state. Using the mass and charge state values in Table~\ref{tab:charge_states}, we consider all combinations of ionic ratios and calculate their \Sval--value. The corresponding values are shown in Table~\ref{tab:s_values}.

\subsection{2012 August 31 event}

\begin{figure*}[htb]
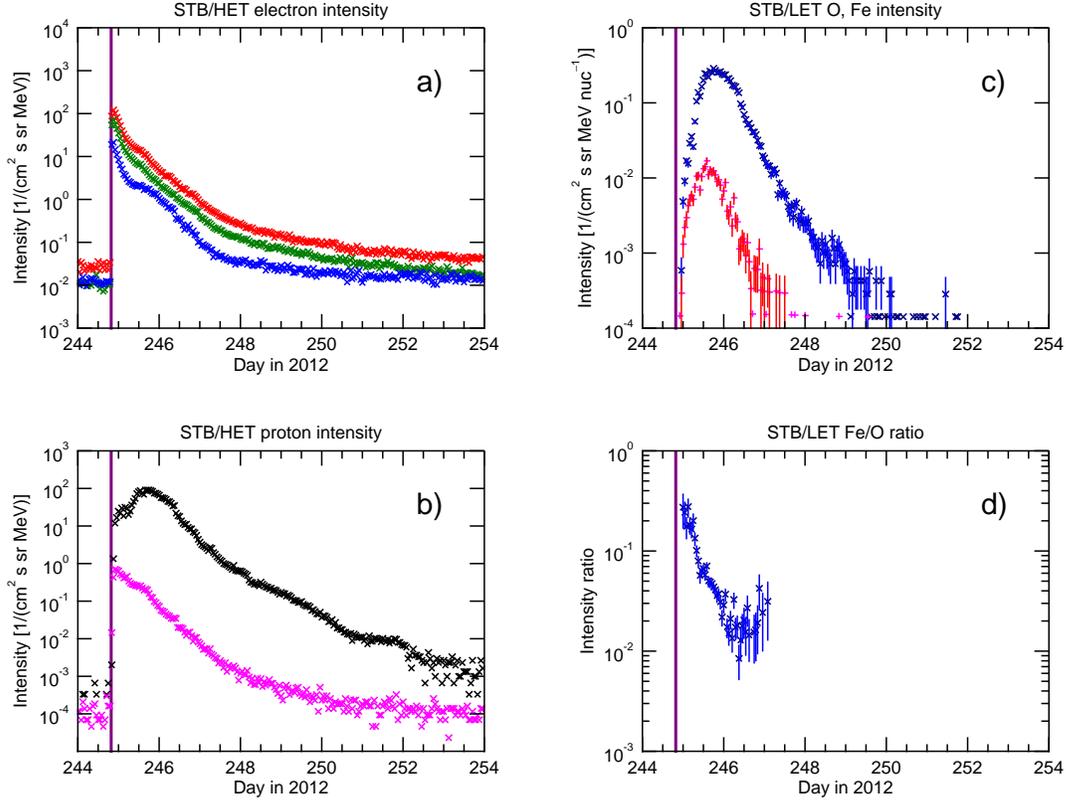


\centering
   \includegraphics[width=0.96\textwidth]{{{sep_apj15_STB_HET_2012_DOY_244-254_h}}}

\caption{SEP event of 2012 August 31 (event \#1): a) relativistic electron intensity (energy channels: 0.7--1.4 (red), 1.4--2.8 (green), 2.8--4.0~MeV (blue)), b) proton intensities at high (40--60~MeV, magenta) and low (13.6--15.1~MeV, black) energy, c) Fe (red) and O (blue) intensity at 4.0--4.5~MeV/nuc, d) Fe/O intensity ratio. Electron and proton intensities were measured by STEREO B/HET, Fe and O intensity by STEREO B/LET. The vertical purple line denotes the start time of the flare. \label{fig:sep_event1}}

\end{figure*}

A filament eruption occurred in the south--eastern region of the solar disk as viewed from Earth on 2012 August 31 (day of year; DOY 244) and launched a coronal mass ejection (CME) with linear speed 1442~km/s\footnote{{\tt http://cdaw.gsfc.nasa.gov/CME\_list/}}. An associated C8 X-ray flare at S16E42 started at 19:45~UT and peaked at 20:43~UT \citep{gallagher2002}. A shock passed the STEREO B spacecraft on September 3 (DOY 247) at 07:11~UT\footnote{{\tt ftp://stereodata.nascom.nasa.gov/pub/ins\_data/ impact/level3/STEREO\_Level3\_Shock.pdf}}. As viewed from Earth the flare and the CME were not particularly strong or fast but they caused a significant particle event at STEREO B with magnetic connection $\Delta\phi = -5\,^{\circ}$. The event (labeled in the following as event \#1) could in many respects be considered an exemplary proton SEP event. Electron, proton, oxygen and iron particle intensity--time profiles and Fe/O ratio are shown in Figure~\ref{fig:sep_event1}. Both the electron and proton intensity profiles, Figure~\ref{fig:sep_event1}, panels a) and b), show a rapid increase followed by a gradual decrease with smooth profile and no significant shock--associated particle intensity component was observed over 10 days. Figures~\ref{fig:sep_event1} c) and d) show oxygen and iron intensities, and the Fe/O ratio, respectively. The intensity of oxygen peaks later than that of iron and the oxygen SEP event lasts longer. The Fe/O ratio is decreasing from values typical of impulsive events to values more than 1 order of magnitude below the average gradual event abundance of 0.134 \citep{reames1995}. This event would be classified as Fe--poor with average Fe/O value over the event $A_{avg} ({\rm Fe/O}) = 0.064$. The Fe/O decrease occurred over $\Delta t_{\rm Fe/O} = 1.25\:\rm days$. After Fe/O reaches its minimum, it increases for the next $\sim\!1 \:\rm day$. Considering the intensity profiles of Fe and O, it is apparent that the increase is a result of the low Fe counts at the limit of the instrument's sensitivity.

\begin{figure*}[t]
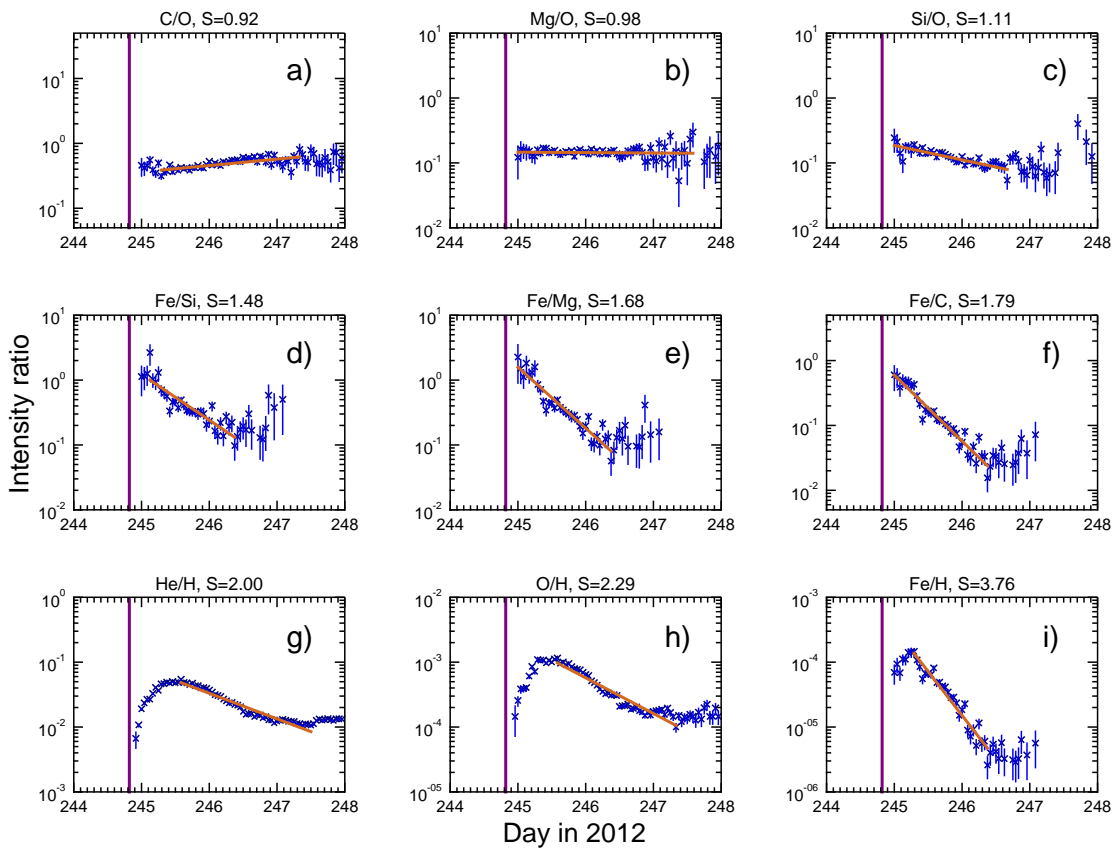


\centering
   \includegraphics[width=0.96\textwidth]{{{sep_apj15_STB_LET_2012_DOY_244-248_h}}}

\caption{A succession of SEP ratios versus time with ascending \Sval values (see Table~\ref{tab:s_values}) for event \#1. SEP ion intensities were measured by STB/LET at 4.0--4.5~MeV/nuc. Ratio data points (blue) in time interval between the maximum and the minimum were fitted to Equation~\ref{eq:log_fit} (plotted in brown), where \Bval is the ratio decay time constant. Ratios with larger \Sval show more temporal evolution, i.e.~lower \Bval. Ratios X/H increase before they start decreasing. All intervals on the vertical axes are scaled equally to 3 orders of magnitude. \label{fig:mq_ratios1}}

\end{figure*}

For the same event, Figure~\ref{fig:mq_ratios1} shows a subset of SEP ratios versus time ordered by increasing \Sval values. The graphs show a 4--day period, during which the temporal evolution of heavy ion ratios takes place. From a qualitative point of view, the time variation of the ratio displays a correlation with \Sval: a ratio shows decrease (increase) in time when $S>1$ ($S<1$). The ratio profiles also show that the slope of temporal variation scales with \Sval, where e.g.~Mg/O shown in Figure~\ref{fig:mq_ratios1} a), a ratio with $S\approx1$, remains almost unchanged over the duration of the SEP event. The O/C ratio with $S=1.09$, Figure~\ref{fig:mq_ratios1} b), shows a slight but steady decrease over time. The rate of decay of Si/O with $S=1.11$, Figure~\ref{fig:mq_ratios1} c), is higher than that of O/C, but the duration of Si/O evolution is also shorter due to the lower relative abundance of silicon in the SEP event. These ratios show far less variation than the ratios with high values of \Sval, i.e.~Fe/Si, Fe/Mg and Fe/C, Figure~\ref{fig:mq_ratios1} panels d), e) and f), all of which have larger \Sval. Ratios of elements with respect to hydrogen, e.g.~He/H, O/H and Fe/H are somewhat anomalous, showing an initial increase followed by a decrease, as can be seen in the last row of Figure~\ref{fig:mq_ratios1} in panels g), h) and i). The decay rate of Fe/H is higher than that of O/H, which is higher than He/H, therefore increasing with increasing $S$ of an SEP ratio. The decay period is shorter for Fe/H than it is for O/H and He/H respectively, due to relative abundances of the elements in the SEP event. The initial increase is observed in all X/H ratios.

\subsubsection{Quantitative analysis}

\begin{figure}[htb]
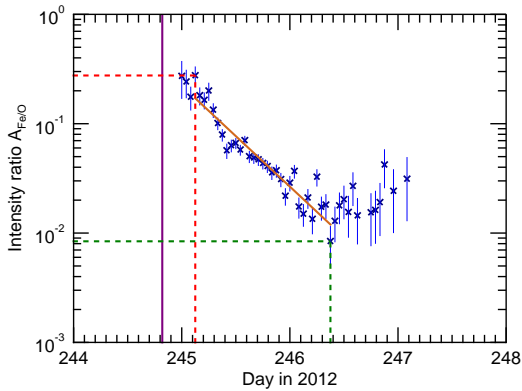

\begin{center}

 \includegraphics[width=0.96\columnwidth]{{{sep_apj15_feo_fit_STB_LET_2012_DOY_244-248_h}}}
\caption{Ratio of Fe/O measured by STEREO B/LET at $4.0-4.5\:\rm MeV/nuc$ with the time of the preceding flare (vertical purple line), maximum (red) and minimum (green). The line of best fit (brown) was fitted to the data in the period between the maximum and minimum. \label{fig:ratio_decrease1}}

\end{center}
\end{figure}

We carried out a quantitative analysis of the temporal variation of heavy ion ratios in the SEP events as follows. We indicate as $A$ the heavy ion ratio, calculated as the ratio of two particle intensity values $I$, e.g.~$A_{\rm Fe/O} = I_{\rm Fe} \big/ I_{\rm O}$. As an example, Figure~\ref{fig:ratio_decrease1} shows the Fe/O ratio in event \#1 at $4.0-4.5\:\rm MeV/nuc$ over a 4--day period. Ratio data points used in further analysis are those that have more than 2 particle counts in a 1--hour time bin for both ion species. An input to the analysis is the time range $\Delta t$ from the start of the flare over which data should be fitted. We marked the first occurring maximum or minimum data point within $\Delta t$ as $A_1$, and the last occurring as $A_2$. When the Fe/O ratio decreases over time, the maximum precedes the minimum and $A_2/A_1<1$, but other heavy ion ratios may show an increase over time, i.e.~$A_2/A_1>1$. All the data points between the maximum and the minimum were fitted using the function 

\begin{equation}
  A = 10^{\alpha + Bt}
  \label{eq:log_fit}
\end{equation}

\noindent where $B$ is the ratio decay time constant in units of $\rm day^{-1}$, and $\alpha$ is a unitless fitting constant. For the example of Figure~\ref{fig:ratio_decrease1} the decay time constant, obtained using Equation~\ref{eq:log_fit}, is $B_{\rm Fe/O}=(-0.92\pm0.06)\:\rm day^{-1}$. In a similar manner, the values of $B$ were obtained for all abundant ratios considered in Table~\ref{tab:s_values}.

\begin{figure}[htb]
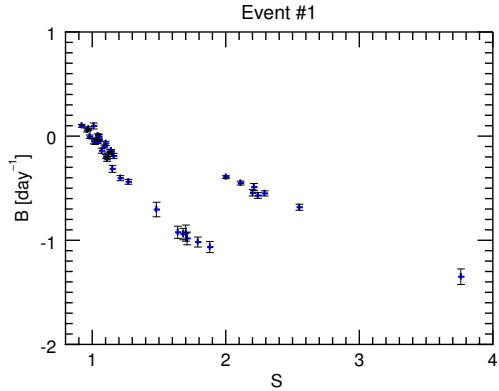

\begin{center}

 \includegraphics[width=.96\columnwidth]{{{sep_ratio_analysis_STB_LET_2012_DOY_244-248_h_coeff2_slope-mq}}}
\caption{Decay time constants $B$ plotted as a function of \Sval for event \#1. The monotonic dependence shows that more decrease is observed in ratios with increasing \Sval. A discontinuity is observed at $S=2.0$ (He/H). \label{fig:b_vs_s1}}

\end{center}
\end{figure}

In Figure~\ref{fig:b_vs_s1} we plot the decay time constant $B$ as a function of \Sval for all SEP ratios in event \#1. The plot shows a quantitative description of the qualitative behaviour seen in Figure~\ref{fig:mq_ratios1}. Between $S=1$ and $S=2$ a monotonic decrease of $B$ with \Sval is observed corresponding to faster decay rates as \Sval increases. There is a discontinuity present at $S=2.0$, the $S$ value for He/H, followed by another monotonic decrease. As can be seen from Table~\ref{tab:s_values}, the ratios with $S\geq2$ are ratios of a heavy ion and hydrogen, X/H.

{ 
The location of the data points along the horizontal axis is influenced by the values of the charge states $Q$ that are used to calculate $S$. We analysed how $S$ values change when different values of $Q_{\rm Fe}$, the $Q$ value that can vary by the largest amount, are considered. The $S$ value for Fe/O in our plot is $S=1.64$. This value changes to $S=2.04$ when $Q_{\rm Fe}=12$ and $S=1.36$ when $Q_{\rm Fe} = 18$. If the Fe charge changed, ionic charges for all the other ions would change too, therefore it is not easy to quantify the effect on the $B$ vs $S$ plot unless charge states for the event for all ions were available. }

\subsection{2006 December 13 event}

Next we considered the event on 2006 December 13 (DOY 347, event \#2), which shows similar heavy ion profiles as event \#1. Event \#2 was caused by an X3.4 flare at S05W23, which started at 02:14~UT and peaked at 02:40~UT. The flare was accompanied by a CME with speed 1774~km/s and an interplanetary shock passed ACE on December 14th (DOY 348) at 14:14~UT. The event was studied in detail by \citet{liu2008} and \citet{rosenvinge2009}. Particle intensity profiles for event \#2 are shown in Figure~\ref{fig:sep_event2}. Event \#2 was followed by another SEP event on December 14 at 22:14~UT, therefore the analysed period was cut off on 2006 December 15 (DOY 349). The SEP event occurred while the particle intensity was elevated from a preceding event, see Figure~\ref{fig:sep_event2} b), but the proton intensity increase due to event \#2 was several orders of magnitude. The pre--event background for heavy ions was less significant than for protons because heavy ion events decay faster. We use the heavy ion data from ACE because the STEREO spacecraft were still performing manoeuvres near the Earth \citep{rosenvinge2009}. At the time of the flare, the magnetic connection of ACE was $\Delta\phi=-17\,^{\circ}$. The Fe/O ratio, Figure~\ref{fig:sep_event2} d), decreases for $\Delta t_{\rm Fe/O} = 1.25\:\rm days$. Fe/O decays to a value of about $2\times10^{-1}$, higher than the value reached in event \#1. This event is Fe--rich with average Fe/O value $A_{avg} ({\rm Fe/O}) = 0.540$. Figure~\ref{fig:mq_ratios2} shows a subset of heavy ion ratios as in Figure~\ref{fig:mq_ratios1}. The ratios show similar temporal evolution as was observed in event \#1, except for the initial increases observed in most heavy ion ratio time profiles, which may be the result of the elevated background from the preceding event.

\begin{figure*}[htb]

\centering
   \includegraphics[width=0.96\textwidth]{{{sep_apj15_SOH_ERN_2006_DOY_347-349_h}}}

\caption{SEP event 2006 December 13 (event \#2): a) relativistic electron intensity (energy channels: 0.7--1.4 (red), 1.4--2.8 (green), 2.8--4.0~MeV (blue)), b) proton intensities at high (40.5--62.2~MeV, magenta) and low (13.8--14.6~MeV, black) energy, c) Fe (10.7--15.8~MeV/nuc, red) and O (10.0--13.1~MeV/nuc, blue) intensity, d) Fe/O intensity ratio. Electron intensities were measured by STEREO B/HET, proton intensities by SOHO/ERNE, Fe and O intensity by ACE/SIS. The vertical purple line denotes the start time of the flare. \label{fig:sep_event2}}

\end{figure*}

\begin{figure*}[t]

\centering
   \includegraphics[width=0.96\textwidth]{{{sep_apj15_ACE_SIS_2006_DOY_347-349_h}}}

\caption{A succession of SEP ratios with ascending \Sval values (see Table~\ref{tab:s_values}) for event \#2. Proton intensities were measured by SOHO/ERNE and heavy ion intensities by ACE/SIS. Energy channels for a pair of ions in each ratio, which were selected to be the closest match, are the following: C/O -- C 6.4--8.6~MeV/nuc, O 7.3--10.0~MeV/nuc; Mg/O -- Mg 8.7--12.2~MeV/nuc, O 10.0--13.1~MeV/nuc; Si/O -- Si 9.2--13.0~MeV/nuc, O 10.0--13.1~MeV/nuc; Fe/Si -- Fe 10.7--15.8~MeV/nuc, Si 9.2--13.0~MeV/nuc; Fe/Mg -- 10.7--15.8~MeV/nuc, Mg 12.2--16.0~MeV/nuc; Fe/C -- Fe 10.7--15.8~MeV/nuc, C 11.2--13.4~MeV/nuc; He/H -- He 3.4--4.7~MeV/nuc, H 3.5--4.1~MeV; O/H -- 7.3--10.0~MeV/nuc, H 8.1--10.1~MeV; Fe/H -- Fe 10.7--15.8~MeV/nuc, H 13.8--14.6~MeV. \label{fig:mq_ratios2}}

\end{figure*}

\begin{figure}[htb]
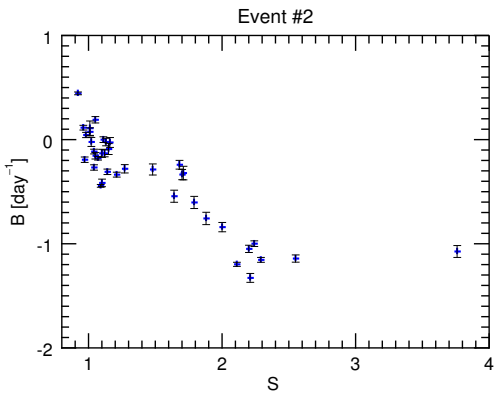


\centering
	\includegraphics[width=0.96\columnwidth]{{{sep_ratio_analysis_ACE_SIS_2006_DOY_347-349_h_coeff2_slope-mq}}}

\caption{Decay time $B$ plotted as a function of \Sval for event \#2. The monotonic dependence as in event \#1 is observed but without the discontinuity. \label{fig:b_vs_s2}}

\end{figure}

We applied the quantitative analysis to the SEP ratios as in event \#1 and plotted the function $B$ vs \Sval as shown in Figure~\ref{fig:b_vs_s2}. The dependence shows a monotonic behaviour but in this case the discontinuity at $S=2.0$ is not present. Instead, the X/H ratios decay at faster rate than other heavy ion ratios. This behaviour is observed because the proton intensity at these energies only changes $\approx\!1$ order of magnitude over the 2--day period (cf.~proton intensity at 13.8--14.6~MeV in Figure~\ref{fig:sep_event2} b)).

\subsection{2014 September 1 event}

Event \#3 originated from an active region behind the eastern limb of the Sun as viewed from Earth on 2014 September 1 (DOY 244) at 11:00~UT. We determined the location of the flare at N14E129 using FITS data from the STEREO B EUVI instrument. \citet{pesce2015} estimated the flare class as X2.1 located at N14E126. The associated CME had speed 1901~km/s. STEREO B, which was the best magnetically connected spacecraft ($\Delta\phi = -25\,^{\circ}$), encountered a passing shock on September 3 (DOY 246) at 07:45~UT\footnote{{\tt ftp://stereodata.nascom.nasa.gov/pub/ins\_data/ impact/level3/STEREO\_Level3\_Shock.pdf}}. The particle intensity profiles are shown in Figure~\ref{fig:sep_event3}. A sudden increase in the oxygen particle intensity of $\approx\!1$ order of magnitude (Figure~\ref{fig:sep_event3} c), blue) was observed $\sim\!13$~hours after the flare started, which is not present in the iron particle intensity (red). This increase in oxygen occurred more than a day before the shock passed the spacecraft. The Fe/O ratio dropped rapidly due to the increase in oxygen intensity and reached its minimum $\Delta t = 1.25$~days after the maximum. The event--averaged Fe/O value is $A_{avg} ({\rm Fe/O}) = 0.089$. Figure~\ref{fig:mq_ratios3} shows a subset of heavy ion ratios. Many heavy ion ratios, including Fe/Si, Fe/Mg, Fe/C and Si/O, show a rapid drop in ratio value followed by a plateau, a similar to the Fe/O time. Apart from the sudden drop, the ratios in event \#3 show qualitatively similar behaviour to events \#1 and \#2 in the ordering of the decreases (increases) by \Sval.

\begin{figure*}[htb]

\centering
   \includegraphics[width=0.96\textwidth]{{{sep_apj15_STB_HET_2014_DOY_244-254_h}}}

\caption{SEP event 2014 September 1 (event \#3): a) relativistic electron intensity (energy channels: 0.7--1.4 (red), 1.4--2.8 (green), 2.8--4.0~MeV (blue)), b) proton intensities at high (40--60~MeV, magenta) and low (13.6--15.1~MeV, black) energy, c) Fe (red) and O (blue) intensity at 4.0--4.5~MeV/nuc, d) Fe/O intensity ratio. Electron and proton intensities were measured by STEREO B/HET, Fe and O intensity by STEREO B/LET. The vertical purple line denotes the start time of the flare. \label{fig:sep_event3}}

\end{figure*}

\begin{figure*}[htb]

\centering
   \includegraphics[width=0.96\textwidth]{{{sep_apj15_STB_LET_2014_DOY_244-248_h}}}

\caption{A succession of SEP ratios with ascending \Sval values (see Table~\ref{tab:s_values}) for event \#3. As in Figure~\ref{fig:mq_ratios1}, the SEP ion intensities were measured by STB/LET at 4.0--4.5~MeV/nuc. The SEP ratio data (blue) are overplotted with the fitted function (brown). \label{fig:mq_ratios3}}

\end{figure*}

\begin{figure}[tbhb]
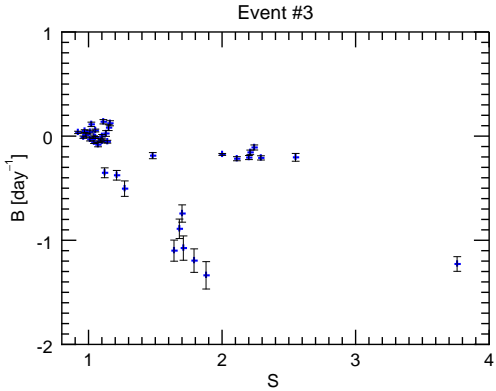


\centering
  \includegraphics[width=0.96\columnwidth]{{{sep_ratio_analysis_STB_LET_2014_DOY_244-248_h_coeff2_slope-mq}}} 

\caption{Decay time $B$ plotted as a function of \Sval for event \#3. The dependence is qualitatively very similar to event \#1. \label{fig:b_vs_s3}}

\end{figure}

We applied the quantitative analysis to the SEP ratios as in event \#1. The values of decay time constant \Bval obtained by fitting in this event largely depend on the length of the interval $\Delta t$, where the data points are fitted. For example, \Bval values obtained in two ratios Fe/Si and Fe/Mg, Figure~\ref{fig:mq_ratios3} panels d) and e), are $B_{\rm Fe/Si} = (-0.19\pm0.03)\:\rm day^{-1}$ and $B_{\rm Fe/Mg} = (-0.89\pm0.09)\:\rm day^{-1}$ but the corresponding fitting interval for Fe/Si ($\Delta t_{\rm Fe/Si} = 2.63\:\rm day$) is more than twice as long as it is for Fe/Mg ($\Delta t_{\rm Fe/Mg} = 1.17\:\rm day$). Such difference can be seen in O/H and Fe/H ratios in Figure~\ref{fig:mq_ratios3} panels h) and i).  We plotted $B$ vs \Sval, shown in Figure~\ref{fig:b_vs_s3}, and the data point at $S=1.48$ corresponding to Fe/Si deviates from the otherwise monotonic dependence in $S \in [0.9,2.0)$. Nevertheless, the obtained dependence qualitatively resembles the \Bval vs \Sval dependence in event \#1, Figure~\ref{fig:b_vs_s1}.

\subsection{2014 February 25 event}

SEP event \#4, which occurred on 2014 February 25 (DOY 56), was caused by an X4.9 flare that erupted at S12E77 and it was accompanied by an exceptionally fast CME with speed 2147~km/s. STEREO B was the best magnetically connected spacecraft with the magnetic footpoint separated from the flare by $\Delta\phi = 32\,^{\circ}$. Figure~\ref{fig:sep_event4} shows intensity profiles for the event at STEREO B. Fe and O intensities were also detected by ACE and STEREO A but these spacecraft were not magnetically well connected \citep{zelina2015_icrc}. As can be seen in Figure~\ref{fig:sep_event4} panel d), the SEP event at STEREO B shows the Fe/O ratio decreased for $\Delta t = 2.75$~days but a significant decrease occurred during the first $\sim\!12$~hours followed by Fe/O remaining relatively unchanged for the rest of the event. This event is Fe--rich with average Fe/O value $A_{avg} ({\rm Fe/O}) = 0.212$. Heavy ion ratios of event \#4 are shown in Figure~\ref{fig:mq_ratios4}. Similarly to the Fe/O ratio, other heavy ion ratios too show little variation over time after $\approx\!1$~day.

\begin{figure*}[htb]

\centering
   \includegraphics[width=0.96\textwidth]{{{sep_apj15_STB_HET_2014_DOY_55-65_h}}}

\caption{SEP event 2014 February 25 (event \#4): a) relativistic electron intensity (energy channels: 0.7--1.4 (red), 1.4--2.8 (green), 2.8--4.0~MeV (blue)), b) proton intensities at high (40--60~MeV, magenta) and low (13.6--15.1~MeV, black) energy, c) Fe (red) and O (blue) intensity at 4.0--4.5~MeV/nuc, d) Fe/O intensity ratio. Electron and proton intensities were measured by STEREO B/HET, Fe and O intensity by STEREO B/LET. The vertical purple line denotes the start time of the flare. \label{fig:sep_event4}}

\end{figure*}

\begin{figure*}[htb]

\centering
   \includegraphics[width=0.96\textwidth]{{{sep_apj15_STB_LET_2014_DOY_56-60_h}}}

\caption{A succession of SEP ratios with ascending \Sval values (see Table~\ref{tab:s_values}) for event \#4. As in Figure~\ref{fig:mq_ratios1}, the SEP ion intensities were measured by STB/LET at 4.0--4.5~MeV/nuc. The SEP ratio data (blue) are overplotted with the fitted function (brown). Much of the temporal variation is observed during the first day of the event. \label{fig:mq_ratios4}}

\end{figure*}

The quantitative analysis was applied to the SEP data in event \#4 in two ways: first the fitting procedure was applied to the entire event (i.e.~over a time range $\Delta t = 4$~days), second it was applied only to the first 12 hours after the start of the flare. The first 12 hours are when a fast decrease (increase) occurs in some ratios. The graphs of $B$ vs \Sval dependence for the two $\Delta t$ values are shown in Figure~\ref{fig:b_vs_s4}. The $B$ vs \Sval graph for $\Delta t = 4$~days, Figure~\ref{fig:b_vs_s4} a), shows the decay time constants are smaller because the initial decrease in Fe/X ratios is averaged over a longer period, over which the ratios remain relatively unchanged. A discontinuity can hardly be observed due to small $B$ values of the ratios. As can be seen in Figure~\ref{fig:mq_ratios4} b), when the fit is carried out over 12 hours, the obtained $B$ values are larger in magnitude than in the previous events. In the latter plot, a discontinuity at $S=2.0$ is observed, as in Figure~\ref{fig:b_vs_s1}.

\begin{figure}[htb]

\centering
	\includegraphics[width=0.96\columnwidth]{{{sep_ratio_analysis_STB_LET_2014_DOY_56-60_h_coeff2_slope-mq}}} \\

\centering
	\includegraphics[width=0.96\columnwidth]{{{sep_ratio_analysis_STB_LET_2014_DOY_55-65_h_coeff3_slope-mq}}}

\caption{Decay time $B$ plotted as a function of \Sval for event \#4. The top panel shows the results of fits using $\Delta t = 4$~days, the bottom one using $\Delta t = 12$~hours. Note the different scaling of the ordinate. A typical size of errorbars ($\pm1\sigma$) is shown in the right bottom corner of the graphs. \label{fig:b_vs_s4} }

\end{figure}

\section{Discussion}
\label{sec:discussion}

We have studied solar energetic particle (SEP) intensities and elemental ratios in 4 SEP events, where Fe/O has been observed to decrease over time, in the $\approx\!4-15\:\rm MeV/nuc$ energy range. We used 1--hour averaged SEP data from energetic particle telescopes onboard ACE, SOHO, STEREO B, and systematically quantified the temporal dependence of abundant SEP ratios. Each of the SEP events was observed by a well--connected spacecraft with a magnetic footpoint within 32\deg of the flare.

We observed that time evolution of heavy ion ratios is a common feature present in all four analysed SEP events, with largest variation in Fe/X ratios, where X indicates an abundant SEP element. We found that some abundance ratios, e.g.~Mg/O, remained relatively unchanged during an SEP event and some ratios, e.g.~Ne/O, even showed an increase over time. This behaviour is ordered by the \Sval value of an SEP ratio, defined as the ratio of \MQ values of the two SEP species in the ratio. Ratios with $S<1$ ($S>1$) exhibit an increase (decrease) over time. We also observed that the slope of a ratio tends to be steeper for ratios with larger \Sval value. Therefore, the temporal evolution of SEP heavy ion ratios shows ordering by \MQ.

We quantitatively examined the ratios listed in Table~\ref{tab:s_values} for the 4 SEP events, and for each event we plotted the values of decay time constant $B$ as a function of \Sval. Each of the obtained plots, Figures~\ref{fig:b_vs_s1}, \ref{fig:b_vs_s2}, \ref{fig:b_vs_s3} and~\ref{fig:b_vs_s4}, shows a monotonic dependence of $B$ vs \Sval in the range $S \in [0.9,2.0)$. The $B$ vs \Sval dependence in 3 out of 4 events, (including Figure~\ref{fig:b_vs_s4} b)), shows a discontinuity at $S=2.0$ which corresponds to He/H. In event \#2, Figure~\ref{fig:b_vs_s2}, the discontinuity at $S=2.0$ is not present, probably because the proton intensity, shown in Figure~\ref{fig:sep_event2} b), only varies by $\sim\!1$ order of magnitude during the analysed period and does not decay significantly. As a result, the X/H ratios in event \#2 decay at much faster rate than in the other events. Event \#3 showed \Bval vs \Sval dependence qualitatively similar to event \#1. The discontinuity was not observed in event \#4, Figure~\ref{fig:b_vs_s4} a), where $\Delta t = 4$~days was much longer that the period of significant temporal variation, i.e.~the first $\sim\!12$~hours after the start of the flare. The \Bval values obtained by fitting in the two time intervals $\Delta t = 4$~days and 12 hours in event \#4 showed ordering by \MQ.

When plotted in logarithmic--linear plot, intensity ratio time data often exhibit time profiles similar to a linear function between their maximum and minimum values. Every SEP ratio had a minimum and a maximum identified independently. The method of finding the maximum and the minimum of a ratio time profile affected the value of the fitted decay constant \Bval, where ratios fitted over a longer time interval showed less average temporal variation, in particular Fe/Si ratio in event \#3 and most of the ratios in event \#4. This is clearly a limitation of the used method, nevertheless, it allowed us to characterise and quantify the observed temporal evolution in SEP ratio time profiles.

{
Some events have more complex structure than the monotonic intensity decrease shown for example by event \#1. More complicated intensity profiles, from which the ratio profiles are derived, can be caused by interplanetary structures affecting the propagation of ions, multiple events or shocks. This makes the choice of the fit interval $\Delta t$ more challenging and different ionic ratios might require different $\Delta t$ values in a single event (see e.g.~event \#3). }

The temporal evolution of heavy ion ratios has previously been interpreted as a signature of a rigidity--dependent acceleration \citep{tylka1999}. In this model, the decrease in the Fe/O ratio would be caused by Fe ions with large \MQ spending less time at the shock during acceleration and being released earlier than O ions.

However, \citet{mason2006} presented data on Fe and O intensity profiles at two energy ranges, where they showed that the decrease of Fe/O over time is likely a result of SEP propagation through the interplanetary medium, in common with other authors (e.g.~\citealt{scholer1978,mason2012,tylka2013}). In 1D propagation, the scattering mean free path $\lambda$ is assumed to depend on \MQ of an SEP ion, e.g.~$\lambda_{\rm Fe} > \lambda_{\rm O}$. The stronger scattering causes a slower propagation of an ion with lower \MQ values to the observer, which can result in temporal dependence of a ratio profile such as ratios observed in this study.

In a recent study, \citet{dalla2016} carried out 3D full--orbit test particle simulations of SEP propagation in a unipolar Parker spiral magnetic field. Fe and O ions injected near the Sun were allowed to propagate in 3D with a rigidity independent scattering mean free path $\lambda=1\:\rm AU$. Their crossings of the 1~AU sphere were counted as a function of time for observers at various locations with respect to the source. The particles experience strong curvature and gradient drifts \citep{dalla2013} which cause transport in a direction perpendicular to the magnetic field, with drift velocities proportional to $M/Q$. For particles at equal energy/nucleon, the ion with larger \MQ propagates across the magnetic field more easily and is able to reach the detector at earlier time. The calculated Fe/O time profiles from the simulations are qualitatively similar to the observations in events \#1--3. Therefore drift as an SEP transport mechanism in 3D could explain the observed temporal variation of heavy ion ratios. In this model, the simulation with the value of mean free path $\lambda=0.1\:\rm AU$ produced similar results as in the case with $\lambda=1\:\rm AU$. Comparing the results corresponding to the two values of $\lambda$, the variation of the mean free path had little effect on the final Fe/O ratio time profiles.

{
Several ionic ratios in our figures show an increase towards the end of the SEP event, e.g.~Fe/O in Figure~\ref{fig:sep_event1} and Fe/Si, Fe/C and Fe/H in Figure~\ref{fig:mq_ratios1}. This kind of behaviour can also be seen in Figure~5 in \citet{tylka2013}, Figure~2 in \citet{zelina2015_icrc} and Figure~3 in \citet{reames1990}. Such an effect could be in some cases a result of low count statistics, however, for some events it does appear to be a real effect, e.g.~events \#1 and \#4 (Figures~\ref{fig:sep_event1} and \ref{fig:sep_event4}, panels d). { Simulations by \citet{dalla2016} show a similar behaviour in the Fe/O time profiles. In that model the increase late in the event results from O decaying at a faster rate than Fe since O occupied a narrower longitudinal extent than Fe, due to a smaller drift. The increases could be caused by passing magnetic structures, e.g.~shocks and interplanetary coronal mass ejections (ICMEs), within which the particle populations, magnetic field vectors and transport conditions may be different from the surrounding environment.} In event \#2, after passing of the shock on 2006 December 14 at 14:14~UT (Figure~\ref{fig:sep_event2} c), the O intensity suddenly starts decreasing at a faster rate than Fe, and the Fe/O value increases as the result. The Fe intensity profile in event \#3 (Figure~\ref{fig:sep_event3} c) has a smooth intensity profile but O intensity in the same event has a more complicated intensity profile with sharp rises and decreases. As a result, the Fe/O ratio (Figure~\ref{fig:sep_event3} d) also has a complicated time profile with increases and decreases. }

In an analysis of time profiles of heavy ion ratios, \citet{mason2012} noted that the He/H ratio showed a decrease only in some of the 17 SEP events while it did not decrease in others. They concluded that the temporal behaviour of protons was different from heavy ion elements. In our analysis, we found that the time profiles of X/H ratios were anomalous compared to other ratios, often showing an increase of the ratio values before the decrease. The decreases in time occurred at lower rate compared with other heavy ion ratios, which resulted in a discontinuity observed in $B$ versus \Sval plots.

Anomalous time profiles of X/H ratios can emerge in a number of scenarios. The ratios could be signatures of differences in SEP acceleration or interplanetary transport between protons and heavier ions. On the other hand, they could be a result of the much higher abundance of protons in the SEP population. At present, the origin of the anomalous nature of X/H time profiles and their slower decay over time remains unknown, and any theory should be able to explain temporal evolution of heavy ion as well as X/H ratios.

The \Bval vs \Sval profiles depend on our assumption of used charge state values. Obtaining values of SEP charge states is challenging and this type of measurement is not routinely carried out for all SEP events. In our analysis, we used charge state values averaged over a number of events by \citet{luhn1985} that have been used previously in similar studies. While these measurements should be representative of typical ionic charge state values, charge states of an SEP ion are known to have different values in separate particle events (e.g.~$Q_{\rm Fe}\approx10-20$) and can depend on the kinetic energy \citep{klecker2006}. {
If event--specific values of $Q$ were used, $S$--values in Table~\ref{tab:s_values} would be modified and this would result in a shift of the data points horizontally in the \Bval vs \Sval plots (Figures~\ref{fig:b_vs_s1}, \ref{fig:b_vs_s2}, \ref{fig:b_vs_s3} and~\ref{fig:b_vs_s4}). }

\section{Conclusions}
\label{sec:conclusions}

In this study, we used SEP data from ACE/SIS, SOHO/ERNE and STEREO/LET and STEREO/HET to quantitatively characterise the temporal evolution of SEP ratios. Our main results are as follows:

\begin{enumerate}

  \item The temporal evolution of heavy ion ratios is ordered by the ratio of mass--to--charge values of the two SEP ions, $S$.

  \item Between $S=0.9$ and $S=2.0$, considering 28 different ionic ratios, we find a clear monotonic behaviour with \Sval, with slopes of $B$ versus \Sval typically given by --0.5~day$^{-1}$ for event \#2, to --1.5~day$^{-1}$ for event \#3.
  
  \item Ratios of heavy ion to hydrogen, X/H, where X is an abundant SEP element, corresponding to $S\geq2.0$, often show an increase before the decrease in their time profiles and decay at slower rates. This anomaly is present in the $B$ versus \Sval plots as a discontinuity where the $B$ values jump to a significantly higher value than would be predicted by a monotonic $B$ vs \Sval dependence.

\end{enumerate}

Our analysis and previously reported observations of heavy ion SEP data suggest that the temporal variation of heavy ion ratios is a common feature of SEP events. This phenomenon has been observed at various instances in the ecliptic (e.g.~\citealt{tylka1999,mason2012,zelina2015_icrc}) and at high heliographic latitudes \citep{tylka2013}.

In recent years, a consensus has emerged that the time evolution of Fe/O and other ionic ratios is an interplanetary transport effect \citep{mason2006,tylka2013}. At the present time both 1D rigidity dependent scattering ($\lambda \propto (M/Q)^{\alpha}$, e.g.~\citealt{mason2012}) and 3D drift associated transport (drift velocity $\propto M/Q$, \citealt{dalla2016}) are possible mechanisms that may explain our observations.

\acknowledgments

We thank the instrument teams of ACE/SIS, STEREO/LET, STEREO/HET and SOHO/ERNE for providing Level 2 data, made publicly available through ACE and STEREO Science Centers, and the Space Research Laboratory at the University of Turku. P.Z.~acknowledges support from the JHI at the University of Central Lancashire through a PhD studentship. S.D.~acknowledges support from the UK Science and Technology Facilities Council (STFC) (grant ST/M00760X/1) and the Leverhulme Trust (grant RPG--2015--094). The work at Caltech was supported by the National Science Foundation grant NSF--1156004, NASA grants NNX13A66G and subcontract 00008864 of NNX15AG09G.

{\it Facilities:} \facility{ACE}, \facility{GOES}, \facility{SOHO}, \facility{STEREO}.

\bibliography{zelina_SEP.bib}

\end{document}